\documentstyle[epsfig]{mn}
\newcommand{\etalc}{et~al.}

\begin{document}

\title[STJ Observations of HU Aqr]
{STJ Observations of the Eclipsing Polar HU Aqr}

\author[C. M. Bridge et al.]
{C.M. Bridge$^1$, Mark Cropper$^1$, Gavin Ramsay$^1$, M.A.C. Perryman$^2$,
J.H.J. de Bruijne$^2$,\and F. Favata$^2$, A. Peacock$^2$, N. Rando$^2$, A.P. Reynolds$^2$\\
$^1$ Mullard Space Science Laboratory, University College London, Holmbury St. Mary, Dorking, Surrey, RH5 6NT\\
$^2$ Astrophysics Division, Space Science Department of ESA, ESTEC, Postbus
299, 2200 AG Noordwijk, The Netherlands
}

\maketitle


\begin{abstract}

We apply an eclipse mapping technique to observations of the eclipsing magnetic
cataclysmic variable HU~Aqr. The observations were made with the S-Cam2
Superconducting Tunnel Junction detector at the WHT in October 2000, providing
high signal-to-noise observations with simultaneous spectral and temporal
resolution. HU~Aqr was in a bright (high accretion) state ($V=14.7$) and the
stream contributes as much to the overall system brightness as the accretion
region on the white dwarf. The stream is modelled assuming accretion is
occuring onto only one pole of the white dwarf. We find enhanced brightness
towards the accretion region from irradiation and interpret enhanced brightness
in the threading region, where the ballistic stream is redirected to follow the
magnetic field lines of the white dwarf, as magnetic heating from the
stream$\--$field interaction, which is consistent with recent theoretical
results. Changes in the stream eclipse profile over one orbital period indicate
that the magnetic heating process is unstable.

\end{abstract}


\begin{figure}
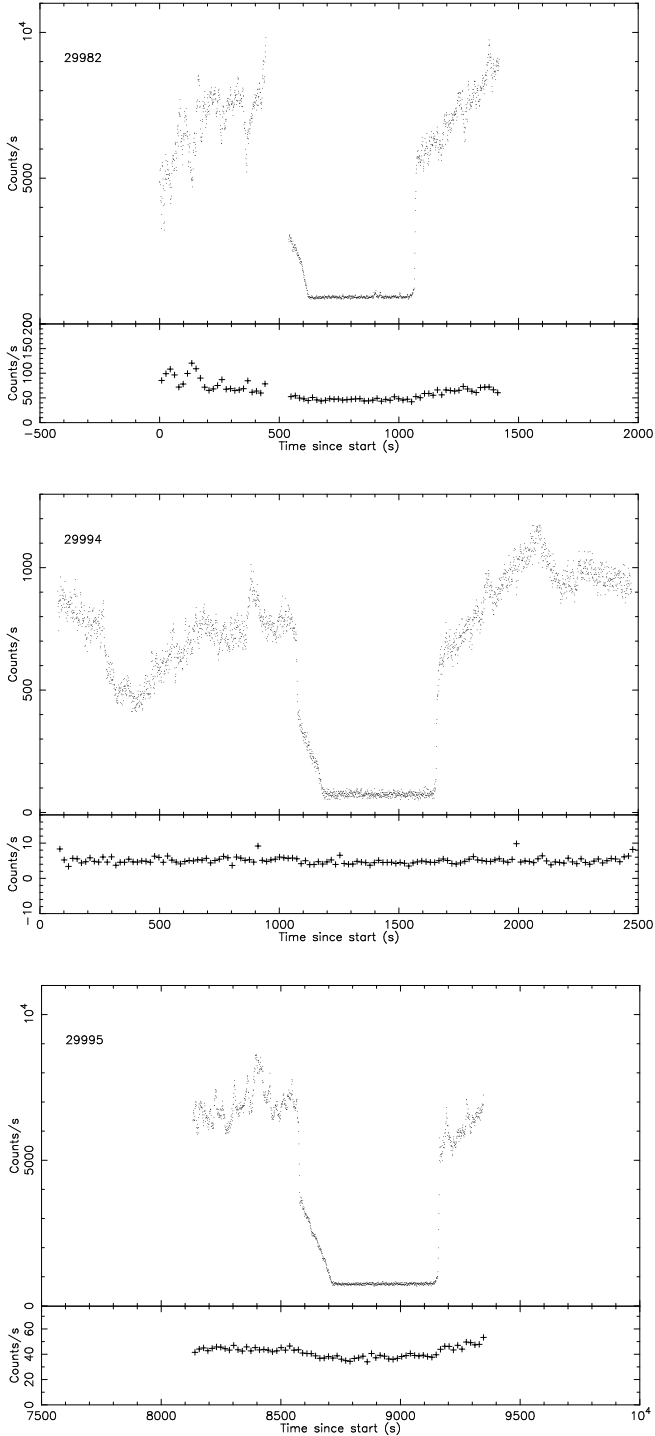

\centerline{\epsfig{file=29982LC.ps,width=6.0cm,angle=-90}}
\vspace{5mm}
\centerline{\epsfig{file=29994LC.ps,width=6.0cm,angle=-90}}
\vspace{5mm}
\centerline{\epsfig{file=29995LC.ps,width=6.0cm,angle=-90}}
\caption{White light curves and sky-backgrounds for cycle 29982 (upper plot),
29994 (middle plot) and 29995 (lower plot). The background is taken from pixels
(1,1) and (6,6), while the light curves are from all 36 pixels. The light
curves are in 1~s time bins (0.00013 phase) and the backgrounds in 18~s time
bins to facilitate direct comparison. The abscissa scales have been adjusted so
the eclipses lie at the same locations. A neutral density filter was used for
cycle 29994, hence the lower count rate and increased noise, and the seeing was
poor for the first part of cycle 29982.}
\label{fig:LCback}
\end{figure}

\section{Introduction}

HU Aqr is a member of the AM Her sub-class of cataclysmic variables, also
called polars because of their highly polarised optical emission. Polars are
characterised by a white dwarf primary with a strong magnetic field
($\sim10-240$ MG) and no accretion disc (see Cropper 1990 and Warner 1995 for
reviews). The secondary is a main sequence star which fills its Roche lobe,
resulting in an overflow of material from the inner Lagrangian point to form an
accretion stream. The stream is initially on a ballistic freefall trajectory,
until at some radius the magnetic pressure of the white dwarf magnetic field is
approximately equal to the ram pressure of the stream. At this point the stream
is threaded onto the field lines of the white dwarf and accretes onto a small
area near one or both of the magnetic poles.

HU~Aqr is an eclipsing polar with a period of $\sim$~$125$~mins, lying just
below the period gap (see e.g. Howell, Nelson \& Rappaport 2001). Harrop-Allin,
Hakala \& Cropper (1999b) used the eclipse of the white dwarf primary by the
secondary to establish the distribution of brightness along the stream given a
number of parameters defining both the geometry of the system and the physical
parameters of the model. The stream emission was calculated using model light
curves which are optimized by a genetic algorithm, based on a method first
employed on photometric eclipse profiles of HU Aqr by Hakala (1995). The
technique was developed and improved by Harrop-Allin \etalc\ (1999a) and
applied to observations of HU~Aqr in both a high and low accretion state (high-
and low-mass transfer; Harrop-Allin \etalc\ 1999b and Harrop-Allin, Potter \&
Cropper 2001).

This method of eclipse mapping has also been applied to emission-lines (Sohl \&
Wynn 1999; Vrielmann \& Schwope 2001; Kube, G\"{a}nsicke and Beurmann
2000). Kube \etalc\ (2000) used HST FOS spectra of (another polar) UZ For to
map the accretion stream line emission at C{\sc iv}\ $\lambda$\ 1550\AA, and
Vrielmann \& Schwope (2001) used the emission line light curves of H$\beta$,
H$\gamma$ and He\ {\sc ii}\ $\lambda$4686\AA\ of HU Aqr. Both these
line-emission methods use a three-dimensional stream so data from the entire
orbit of the system is fitted and the model can reproduce the out-of-eclipse
features, such as the pre-eclipse dips (Watson 1995). They can also reproduce
the effect of differing brightness distributions between the irradiated and
un-irradiated faces of the stream, and this effect is not reproduced in the
model of Harrop-Allin \etalc\ (1999a and 1999b). However, the Harrop-Allin
\etalc\ technique has the benefit of being sensitive to the total emission from
the stream (not just the line emission). The technique also has the advantage
of access to high signal-to-noise ratio data, which is crucial in producing
robust fits.

The light curves presented here were obtained using the S-Cam2 Superconducting
Tunnel Junction (STJ) camera, developed by the ESA Astrophysics Division at
ESTEC. The camera is the second prototype of a new generation of detectors that
record the energy as well as the position and time of arrival (to within
$\sim5\ \mu$s for this particular detector) of the incident photons (Rando
\etalc\ 2000). The application of STJs to optical photon counting was first
proposed by Perryman, Foden \& Peacock (1993), and has since been applied to
observations of the Crab Pulsar (Perryman \etalc\ 1999), the magnetic
cataclysmic variable UZ For (Perryman \etalc\ 2001) and to quasar spectroscopy
(de Bruijne \etalc\ 2002).

The detector itself consists of a liquid helium cooled array of $6\times 6$
pixels, each being $25\times 25$~$\mu$m$^2$, corresponding to $0.6 \times 0.6$
arcsec$^{2}$ per pixel and a field-of-view of about $4 \times 4$ arcsec$^2$. The
pixels are sandwiches of superconducting tantalum with a thin insulating layer
in the middle and the whole device is cooled well below the superconductor's
critical temperature (about 0.1T$_{c}$). An incident photon then perturbs the
device equilibrium and as the energy gap between the ground state and the
excited state is only a few meV, a large number of free electrons is created by
each photon, this number being proportional to the photon energy. This is in
contrast to normal optical CCD semiconductors where the band gap is $\sim1$~eV,
and photon absorption results in typically only one free electron being
created.

The S-Cam2 instrument is particularly suited to observations of cataclysmic
variables. The high-time resolution and simultaneous observations of spectral and intensity variations is ideal for eclipsing systems with orbital
periods of the order of those in cataclysmic variables. The intensity
variations over the rapid ingress and egress can be probed directly, and the
rapid variations of the intensity of the accretion stream can be used to
provide information on the possible mechanisms of emission along the stream.

\section{Observations}

\begin{table*}
\caption{Summary of observations of HU Aqr, where cycle number refers to the ephemeris of Schwope \etalc\ (2001).}
\begin{center}
\begin{tabular}{cclrrl}
\hline
Cycle  &Date	    &Start time &Observation & Egress time    & Remarks\\
number &UTC         &UTC        & length (s) & BJD (TDB)           &\\
\hline
29982  & 2000 Oct 2 & 21:11     & 1420       &                & Data gap; poor seeing\\
29983  & 2000 Oct 2 & 23:11     & 1563       &                & Poor seeing;
data not used\\
29993  & 2000 Oct 3 & 20:13     & 831        &                & Truncated ingress\\
29994  & 2000 Oct 3 & 22:02:42  & 2400       & 2451821.441021 & ND1 filter\\
29995  & 2000 Oct 4 & 00:17:00  & 1216       & 2451821.527841 & Truncated egress\\
\hline
BD+28 4211 & 2000 Oct 2 & 23:46    &       &                  &Standard; ND2 filter; poor seeing\\ 
\hline
\end{tabular}
\end{center}
\label{tab:obslog}
\end{table*}

S-Cam2 was mounted at the Nasmyth focus of the William Herschel
Telescope during 2000 October 2/3 and 3/4. Table~\ref{tab:obslog} gives the
cycle numbers (relative to the ephemeris of Schwope \etalc\ 2001), start time
in UTC, time of mid egress and the observation length in seconds.

A total of five eclipses of HU Aqr were recorded. The ingress was missed for
eclipse 29982 (ephemeris of Schwope~\etalc\ 2001), and only just recorded for
29993. The relatively high count rate of HU Aqr caused some difficulties for
the data acquisition system, limiting the duration of the runs. The first three
eclipses exceeded the data acquisition system limits, resulting in loss of
absolute time reference. Eclipse 29994 was therefore taken with a neutral
density filter with a throughput reduction factor of 10. This is the most
complete of the eclipses, but suffers from reduced signal-to-noise ratio. The
data gap in cycle 29982 was caused by the problems with the data acquisition
system noted above. The seeing was in the range 1 to 1.5 arcsec, except at the
beginning of the run for cycle 29982 and throughout cycle 29983, when it was 2
arcsec, which led to some spillage of light from the $6 \times 6$ array. We
have therefore not used those data.

Observations of the spectrophotometric flux standard BD+28 4211 were taken to
calibrate the data. They were made through a neutral density filter with
attenuation factor 100 and also suffered from poor seeing. 

\section{Data reduction}

The data reduction process employs specific pipeline processing of the S-Cam2
data (de Bruijne \etalc\ 2001), based on the \texttt{FTOOLS} suite of software
(Blackburn 1995) and a full description is given by Perryman \etalc\ (2001)
for their observations of UZ For.

Once reduced, the data can be split {\it a posteriori\/} into different energy
bands (wavelength ranges). The intrinsic STJ resolution is such that
$\lambda$/$\Delta\lambda\sim 9$, but here we use only three bands to represent
`red', `yellow' and `blue' light. The energy bands are selected so that roughly
equal numbers of events occur in each. The wavelength range is
limited by the atmosphere at short wavelengths and the use of optics to
suppress infrared photons at long wavelengths, and corresponds to approximately
340$\--$680~nm. The wavelength ranges for the blue, yellow and red bands here
are 340$\--$470~nm, 470$\--$550~nm and 550$\--$680~nm.

As part of the pipeline processing, the data were flat-field corrected using a
single map derived from sky observations. The data were then corrected for
atmospheric extinction using the standard La Palma tables of extinction values
as a function of wavelength and air mass, and finally sky-background
subtracted. The pipeline processing is applied to a light curve in arbitrary
time bins of 1~s. The extraction of the background-subtracted light curves used
the entire $6\times 6$ array for the object and the two corner pixels (1,1) and
(6,6) for the background. The background light level is then taken as the mean
level during eclipse and this is subtracted from the source light curve.

Figure~\ref{fig:LCback} shows the white light curves for eclipses 29982, 29994
and 29995 before subtraction of the background, together with the backgrounds
taken from the two corner pixels (1,1) and (6,6). 

Once the data had been calibrated and reduced, we folded the data on the
orbital period. The linear ephemeris of Schwope~\etalc\ (2001) was used, which
defines inferior conjunction of the secondary as $\phi=1.0$. The UTC times are
transformed to TDB (at the solar system barycentre, i.e. including light travel
times). For cycles 29982 and 29993, where there is no absolute time reference,
the egress is aligned to be at the same phase as in cycles 29994 and 29995.

\begin{figure*}
\begin{center}
\epsfig{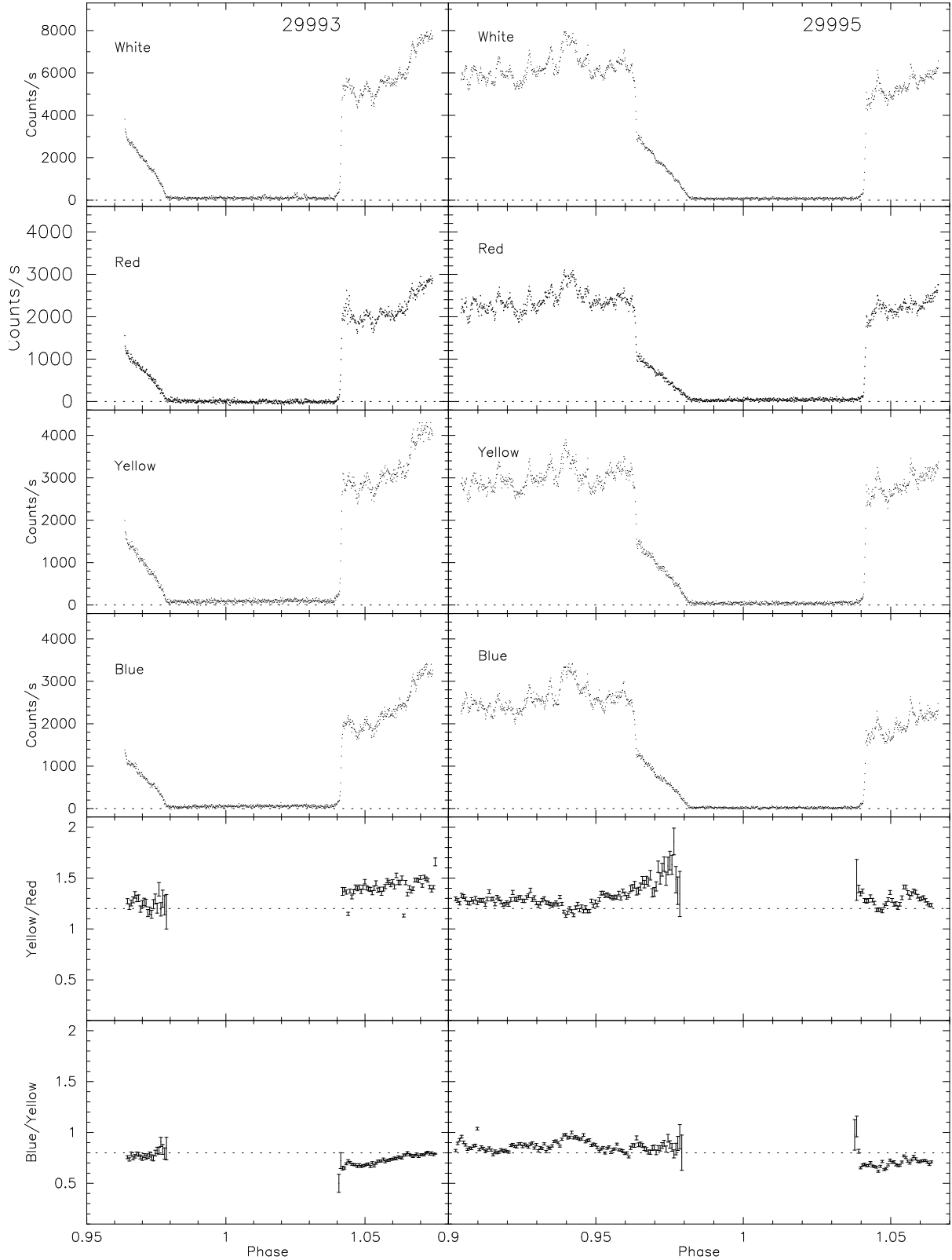}
\end{center}
\caption{The sky-subtracted phase folded light curves for eclipses 29993 (left)
and 29995 (right), representing from the top white, red, yellow and
blue light, with a resolution of 1~s, or 0.00013 phase. The colour ratios are
binned in 5~s bins and the contribution from the secondary during total eclipse
has been removed. Points with errors over 0.5 in the colour ratio plots have
been omitted for clarity -- these occur during complete eclipse when there is
effectively zero count rate after subtraction of the secondary contribution. To
facilitate comparison, dotted lines have been placed at the arbitrary levels of
 0.8 and 1.2 in the blue/yellow and yellow/red ratios respectively.}
\label{fig:huaqr04colours}
\end{figure*}

\begin{figure*}
\begin{center}
\epsfig{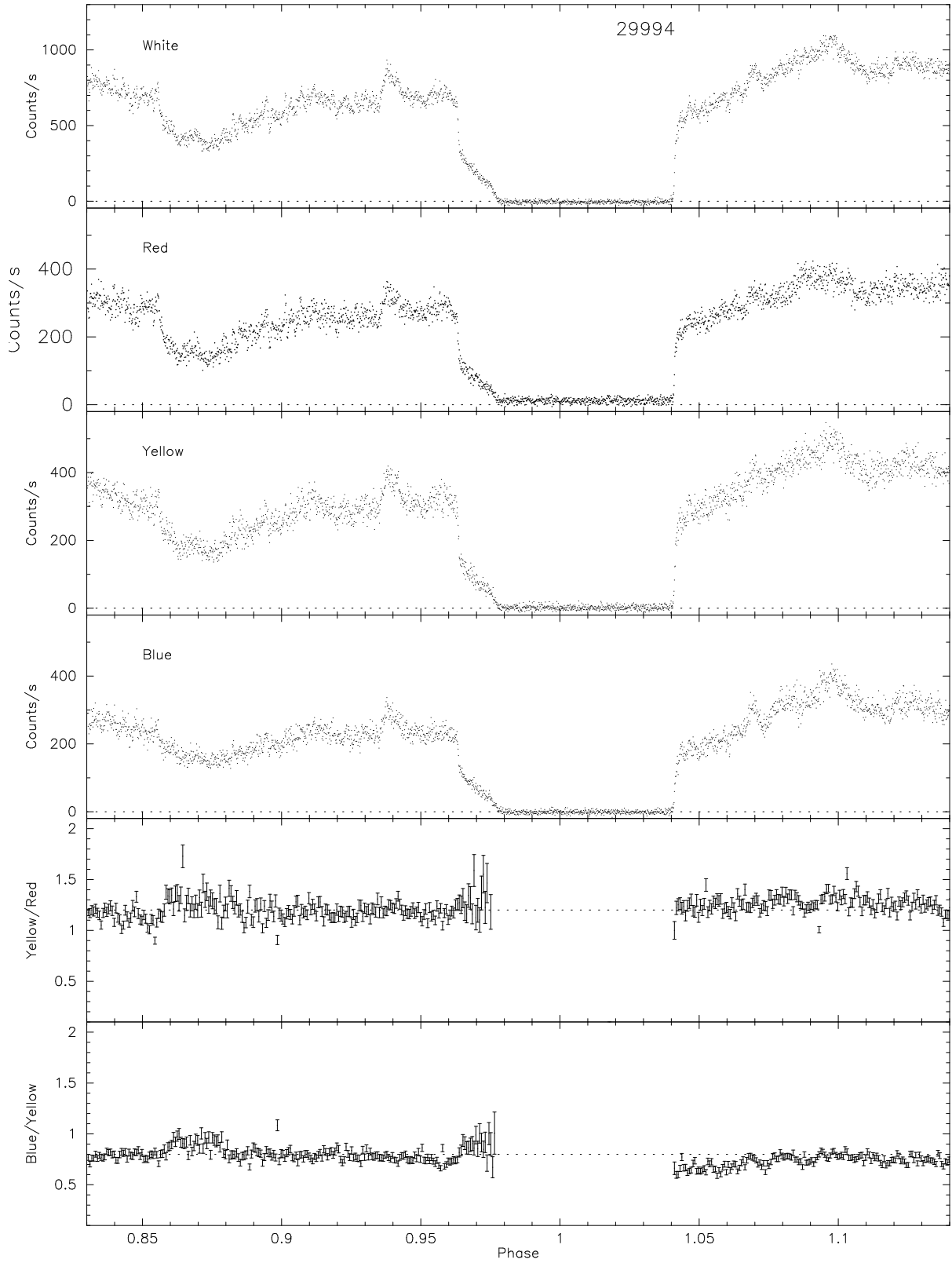}
\end{center}
\caption{As for Figure~\ref{fig:huaqr04colours} but for eclipse 29994. The
countrate is lower by a factor of 10 because of the neutral density filter.}
\label{fig:huaqr03colours}
\end{figure*}

Figures~\ref{fig:huaqr04colours} and \ref{fig:huaqr03colours} show the white,
red, yellow and blue light curves, with the colour ratios yellow/red and
blue/yellow, for cycles 29993, 29994 and 29995. The secondary star has been
subtracted from the colour ratios (but not the light curves) so that the change
in the ratios is due wholly to the stream, after accretion region ingress.

We extracted intervals of good seeing from the standard star observation to
determine the countrate in the yellow band, corresponding most closely to the
$V$ band. This gave a zero point magnitude for 1 count/s in this band of 24.0,
taking into account the (assumed) factor of 100 from the neutral density
filter. The corresponding maximum brightness at $\phi=1.1$ in Figure
\ref{fig:huaqr03colours} is $V=14.7$. This indicates that HU Aqr was in a high
accretion state at the time of these observations (see Schwope \etalc\ 2001).

\section{The Light Curves}

The light curves in Figures~\ref{fig:huaqr04colours}
and~\ref{fig:huaqr03colours} show a number of features that are characteristic
of an eclipsing polar system. The most prominent feature is the eclipse itself,
which starts with the limb of the secondary star eclipsing the accretion
stream. At about $\phi$=0.964 the bright accretion region on the white dwarf is
eclipsed in a few seconds (the white dwarf is also eclipsed around these
phases, but is much fainter). After this only the secondary star is visible,
and the sequence is then approximately reversed on egress.

\subsection{Pre-eclipse dip}

Prior to the eclipse of the white dwarf, we observe a pre-eclipse
dip. Figure~\ref{fig:huaqr03colours} shows this dip centred at $\phi\approx
0.872$. The dip is caused by the eclipse of the accretion region on the white
dwarf by the magnetically confined section of the accretion
stream. Consequently the phase of the centre of the dip is directly related to
the azimuth of the coupling region, where the stream is threaded onto the field
lines of the white dwarf. The centre of the dip in eclipse 29994 corresponds to
an azimuthal angle of 46$^{\circ}$ which is within the range of values found by
Harrop-Allin \etalc\ (1999b) of 42$^{\circ}$ to 48$^{\circ}$ for their light
curves.

Schwope \etalc\ (2001) investigated the correlation between the phase of the
dip and the soft X-ray luminosity, and concluded that the dip ingress occurs
earlier in phase when the system is brighter. The ingress of the dip at
$\phi\approx 0.856$ indicates that the soft X-ray luminosity is at the higher
end of the range, as expected for the bright state of the system.

The dip is visible only in cycle 29994, the other observations being too short
to cover the relevant phases. We cannot therefore relate its movement between
successive eclipses to the change in stream eclipse profile as discussed in
Section~\ref{sec:streameclipsevariations}.

\subsection{Accretion region eclipse}
\label{sec:accretionregioneclipse}

The soft X-ray data in Schwope \etalc\ (2001) suggest that there is only one
accreting pole. It is possible that there is a second accreting pole, as
discussed in Harrop-Allin~\etalc\ (1999b). However, there is no evidence for
such a pole in our eclipse profiles, so for the purpose of our modelling we
assume accretion onto one pole.

\begin{figure}
\centerline{\epsfig{file=spotdiff.ps,width=6.0cm,angle=-90}}
\centerline{\epsfig{file=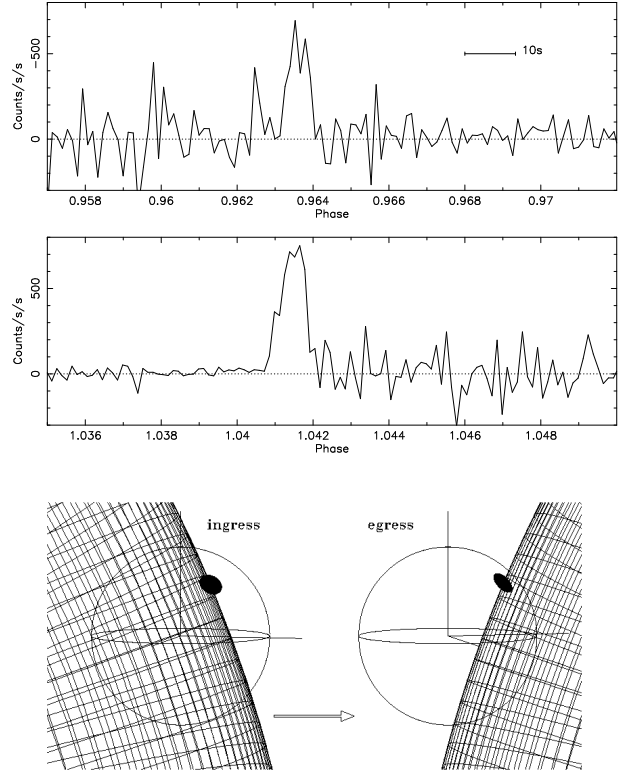,width=5.5cm,angle=-90}}
\caption{(Top) The accretion spot profile on ingress (from cycle 29995) and
egress (the average of cycles 29982, 29993 and 29995). The profiles are
constructed by subtracting successive intensities over 1~s time intervals. The
ordinate of the upper plot is inverted to facilitate comparison. (Bottom) The
view of the eclipsing limb at accretion region mid-ingress and mid-egress. The
accretion region subtending $15^\circ$ is drawn to scale. The radius of the
white dwarf is assumed to be 0.017$a$, where $a$ is the binary separation.}
\label{fig:spot}
\end{figure}

Around $\phi\approx 0.964$ the accretion region has been completely eclipsed
and the accretion stream is thereafter the dominant source of the observed
brightness with a small contribution in the red band from the secondary. We
show in Figure~\ref{fig:spot} the accretion spot profile on ingress and
egress. The profile is constructed by subtracting successive intensities over
1~s time intervals (0.00013 phase) and so shows the rate of change of the light
curves intensity. The profile on egress is the mean of cycles 29982, 29993 and
29995 and is therefore more reliable than the ingress profile, which uses only
cycle 29995. The profile may be asymmetrical, with the leading edge
(corresponding to later phases) of the cyclotron emitting region being brighter
than the trailing edge. The accretion spot ingresses last for $\sim 7$~s and
the egresses are $\sim 8$~s. This is unchanged from the durations measured by
Case (1996) in the data used by Harrop-Allin \etalc\ (1999b) and should be
compared to the duration of $1.3$~s for the soft X-ray eclipses (Schwope
\etalc\ 2001), so that the extent of the cyclotron emitting region is $\sim
15^\circ$ in the optical compared to $\sim 3^\circ$ in soft X-rays. From soft
X-ray absorption modelling, Schwope \etalc\ (2001) suggest that the threading
region is extended in azimuth, but the soft X-ray emission originates over only
a small range of azimuths: the larger region seen in the optical is therefore
likely to be the result of this extension in azimuthal threading. The cyclotron
emitting region is significantly more extended than that measured for UZ For,
where Perryman \etalc\ (2001) found both regions extended over $3^{\circ}\pm
1^{\circ}$.

In those cycles where we have absolute time information (29994 and 29995) the
egress takes place at $\phi = 1.0413$ in both cycles, 10~s later than predicted
by the linear ephemeris. This indicates that the observed-minus-calculated
(O-C) residuals in Schwope \etalc\ (2001) Figure 4 are continuing to
increase. In case there are systematic differences between the optical and soft
X-ray egress times, we rephased the Harrop-Allin (1999) data in their Figure 1
on the new ephemeris, including the UTC-TDB time correction of 59~s appropriate
to the Schwope \etalc\ (1997) ephemeris epoch. The data were taken on cycles
$\sim1322$ in Figure 4 of Schwope \etalc\ (2001), and the O-C of $-1.6$~s is in
agreement to within $\sim 1$~s with those determined from the soft X-ray
data. The continuing increase in O-C residuals is an indication that the period
in the linear ephemeris of Schwope \etalc\ (2001) is slightly too short, and is
in the opposite sense to that expected from their quadratic ephemeris which
predicts a decrease in the O-C residuals. A period of 0.0868204212~days would
eliminate the 10~s residuals we observe.

The duration of the eclipse from the centre of the accretion region ingress to
the centre of its egress is $\Delta\phi=0.0778\pm 0.0002$ in cycles 29993,
29994 and 29995. This is the same as that found by Harrop-Allin \etalc\ (1999b)
of $\Delta\phi=0.0779\pm 0.0002$, and Schwope \etalc\ (1997) of
$\Delta\phi=0.0779$. The lack of variation in the duration of the eclipse is
remarkable and implies that the accretion spot is at the same latitude for
these observations. On the other hand, this is shorter by 6~s than the
$0.0782-0.0792$ modelled by Schwope \etalc\ (2001) from the soft X-ray
eclipses, where the ingress was uncertain due to the low countrate and
absorption.

For a particular mass ratio, the width of the eclipse determines the
inclination of the system. Using $q = M_2 / M_1 = 0.25$ (Schwope \etalc\ 2001),
the inclination is $85.0^\circ$ (assuming $M_1=0.9 M_{\odot}$ and
$\beta=40^\circ$, $\zeta=50^\circ$, where $M_1$ is the mass of the white dwarf
and $\beta, \zeta$ are the magnetic longitude and colatitude of the accretion
region. These correct for the location of the accretion spot on the surface of
the primary; see below).

\subsection{Stream eclipse variations}
\label{sec:streameclipsevariations}

Complete ingress of the stream occurs at $\phi=0.982, 0.979, 0.977$ and $0.982$
for eclipses 29982, 29993, 29994 and 29995 respectively. During the period of
complete eclipse the secondary is the only contributor to the light curve and
provides a constant contribution which is greatest in the red band, as
expected, with no contribution in the blue band.

\begin{figure}
\centerline{\epsfig{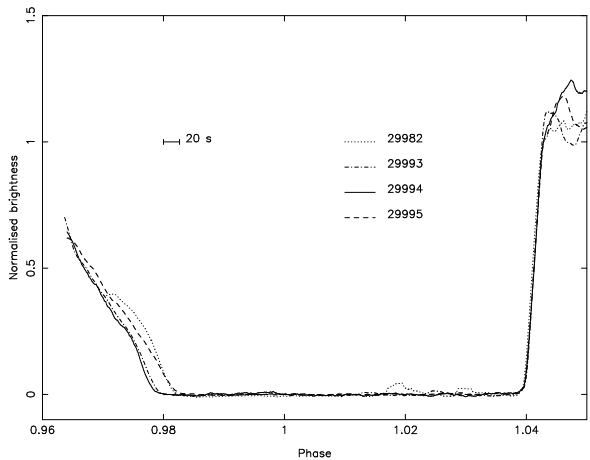}}
\caption{The ingress and egress of eclipses 29982 (dotted), 29993 (dash-dot),
29994 (solid) and 29995 (dashed) smoothed over a running average of 20~s. The
ordinates are normalised to 1 at the end of spot egress, and to 0 at total
eclipse.}
\label{fig:inoutsmooth}
\end{figure}

Figure~\ref{fig:inoutsmooth} shows more clearly that the observations reach
total stream eclipse at different phases even though the spot
ingresses/egresses both occur at the same time. This rapid change in the
accretion flow$\--$magnetic field interaction between successive eclipses was
first observed by Glenn \etalc\ (1994) who found a difference in the time it
took for the stream to be completely eclipsed of more than a minute between
their two successive eclipses.

The shape of the accretion stream eclipse profile in the observed light curves
gives a qualitative prediction of where brightness enhancements can be expected
in the accretion stream. When the stream is brighter at later phases in the
stream eclipse, as in cycles 29982 and 29995 (Figure~\ref{fig:inoutsmooth}),
there must be more emission in the threading region, as this is the only part
of the stream still visible. For the same reason, if the final stage of the
stream eclipse takes place at later phases, the threading region must have
moved: thus in cycles 29982 and 29995, either the accretion stream is brought
further out from the line of centres during the ballistic part of the
trajectory, or the magnetically channeled stream rises further out of the
orbital plane (we address this in Section~\ref{sec:mapoverview} below). This
could be caused either by a small change in the location of the accretion spot
on the primary surface, moving the accretion spot further from the line of
centres, or perhaps a change in the velocity of the stream as it leaves the
L$_{1}$ point. Likewise, if the accretion stream is fainter near the white
dwarf, the system brightness will be fainter immediately after accretion region
ingress. This may be the case for cycle 29995 (Figure~\ref{fig:inoutsmooth}).

\subsection{Colour variations}
\label{sec:colourvariations}

For cycle 29995 there is a rise in the yellow/red ratio during the stream
ingress (Figure~\ref{fig:huaqr04colours}), so that successively more red light
is blocked from the stream as the eclipse progresses. This implies that the
last part of the stream to be eclipsed, near the threading region, emits
relatively less at longer wavelengths than the rest of the stream. This rise is
not evident in cycle 29993, but maybe present at a reduced level in cycle 29994
(Figure~\ref{fig:huaqr03colours}). This trend is consistent with a threading
region which is increasingly hotter in the three successive eclipses. At the
same time as the large yellow/red ratio change in cycle 29995 there is no
change in the blue/yellow ratio.

After eclipse when the stream close to the primary is uncovered first, the
different yellow/red ratios in cycles 29993 and 29995 are consistent with a
hotter stream near the white dwarf in eclipse 29993 than in 29995. An increase
in the blue/yellow ratio after eclipse is seen in all three cycles (29993,
29994 and 29995) and also implies that the magnetically confined stream material is
becoming hotter towards the threading region.

The pre-eclipse dip around $\phi\approx 0.87$ in cycle 29994
(Figure~\ref{fig:huaqr03colours}) is also slightly bluer than at earlier or
later phases. This was also the case in the light curves of
Harrop-Allin~\etalc\ (1999b).

\section{Indirect imaging}
\label{sec:mapoverview}

\subsection{Methodology}

We have inferred from a qualitative analysis of the light curves that the
trajectory and/or intensity of the stream varies over successive orbital
periods, and we can identify brightness enhancements along the stream
length. However, it is more difficult to infer the location of the brightness
enhancements without a model for the stream trajectory. In particular, at
stream ingress there is no information on which parts of the stream are
eclipsed at which phases. We can use an eclipse mapping method to create a
model accretion stream by placing model stream points along a predetermined
trajectory, consisting of a ballistic part from the L$_{1}$ point, coupling to
a magnetically confined section where the material is threaded by the field
lines of the white dwarf. The distance from the white dwarf where this
transition occurs is the threading radius $R_{\mu}$, and is a user input to the
model. The model system is rotated while the Roche lobe filling secondary
eclipses the stream and white dwarf, with the brightness of the visible model
stream points at each phase summed to form a model light curve. The resultant
light curve is optimized with the genetic algorithm (GA) evolving the best fit
light curve.

The `goodness of fit' of a model light curve is measured with a `fitness
function' (see Harrop-Allin \etalc\ 1999a for a full description). The GA
adjusts the brightness points along the stream in an attempt to minimise this
function, which consists of a $\chi^2$ term and a maximum entropy term, which
ensures the problem is not under-constrained. After the final stages of the GA,
a more conventional line-minimisation routine (Powell's method) is used to
reach the final minimum of the solution.

The accretion stream itself is taken to be physically thin in that it does not
eclipse the primary, and so features such as the pre-eclipse dip (e.g. Watson
1995) are not reproduced by the model. However, the brightness contribution of
each stream point is taken as the sine of the angle between the line of sight
and the tangent to the stream at that point, essentially a projection effect,
thus optically thick. We also exclude the white dwarf from the model, although
the egress of the white dwarf can be seen in the light curves immediately
before the egress of the accretion region.

The exact method we apply for the modelling is different to that of
Harrop-Allin~\etalc\ (1999b, 2001) in that we are not applying the technique to
the complete eclipse light curves, nor can we interpret the results in the same
manner. There are a number of reasons for this. Firstly, and importantly, our
observations are all truncated, either in the ingress or the egress of the
accretion stream. For observations where the egress is truncated (cycle 29995),
the model has difficulty breaking the ambiguity which exists when assigning
brightness to points eclipsed in the same phase interval, i.e. the brightness
can be either in the magnetically confined region or the ballistic. For
observations where the pre-eclipse light curve is missing, we lack information
on the accretion region, and parts of the stream towards the secondary (cycle
29993). A further difficulty lies in determining the nature of the variable
emission from the accretion region. This is particularly important at egress,
where the brightness of the spot can greatly affect the brightness distribution
towards the white dwarf. After the egress of the accretion region the light
curve consists of a variable component, plus the stream, which changes rapidly
over the course of the observation. We have therefore chosen to use a
restricted modelling technique, which relies on the ingress of the accretion
stream alone. This means that we have truncated the light curves, removing all
phases up to and including the ingress of the accretion region, and those after
$\phi=1.0$. This is the only part of the light curve unaffected by the variable
emission from the accretion region. We still have the problem of the ambiguity
of the stream points, but we can make some progress in interpreting our results
by referring back to the light curves and the colour ratios.

\subsection{Fixed parameters and $R_{\mu}$}

A number of parameters is required to produce a model light curve and reproduce
the stream brightness distribution. These fall into two types: physical
parameters such as the masses of the two component stars, and geometric
parameters such as the location of the accretion region on the white dwarf (see
Harrop-Allin~\etalc\ 1999a for details). The model is particularly sensitive to
the exact value chosen for the parameter $R_{\mu}$, but we can constrain the
value using the model stream geometry. If the value is too large then the model
needs to assign a large amount of brightness to a few points. Conversely if
$R_{\mu}$ is too small then no emission is assigned by the model to the points
in the threading region. From brightness maps for a range of values of
$R_{\mu}$ we can determine the end point of the ballistic trajectory and so
provide a constraint on the value of $R_{\mu}$. We have found from our model
fits of the light curves used here that the technique depends heavily on the
data being of a sufficiently high signal-to-noise ratio, which is important
because of the sensitivity to the value of $R_{\mu}$. Therefore, in order to
determine the best fit value to use we further restrict the application of our
model technique to cycles 29993 and 29995 (Figure~\ref{fig:huaqr04colours}).

\begin{figure*}
\begin{center}
\epsfig{file=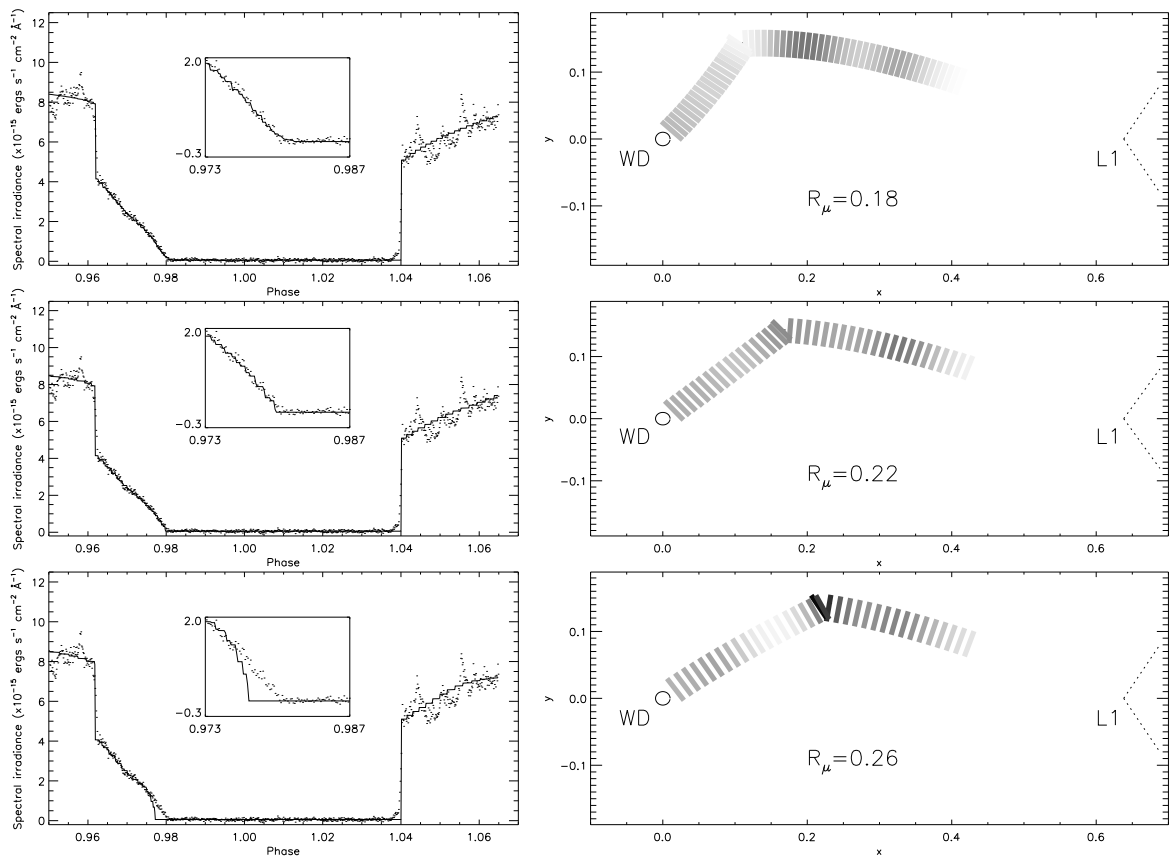,width=12.0cm,angle=90.0}
\end{center}
\caption{Fits to the `blue' light curve of cycle 29995 with different values
for the threading radius $R_{\mu}$ (0.18$a$, 0.22$a$ and 0.26$a$). The stream
brightness maps on the right are gray scale plots with brighter stream emission
being represented as darker. The gradual closing of a `hole' at $R_{\mu}$
illustrates the sensitivity of the model to its location. The insets in the
left hand panels show a close-up view of the fit at the threading region. For
this cycle the value of $R_{\mu}=0.22a$ represents the best stream trajectory.}
\label{fig:rmu29995}
\end{figure*}

Figure~\ref{fig:rmu29995} shows model fits to the `blue' light curve of cycle
29995 for different values of $R_{\mu}$. The stream maps highlight the
dependence of the fits on a correct value for $R_{\mu}$. For a value of
$R_{\mu}=0.18a$ the brightness of the threading region is low. This lack of
emission, or `hole', is caused by $R_{\mu}$ being too small - the data are
incompatible with emission at the end part of the resulting long ballistic
stream. On the other hand, if $R_{\mu}$ is too large, so that the model
ballistic stream is shorter than it is in reality, a pile up of excess
brightness at the threading region is seen, as in the stream map for
$R_{\mu}=0.26$. The inappropriateness of the latter can be deduced from the
poor fit to the region at $\phi=0.98$ (corresponding to the threading region)
for values of $R_{\mu}$ which are too large (insets to
Figure~\ref{fig:rmu29995}).

For all our modelling we used fixed values of $M_1=0.9M_{\odot}$, $q=0.25$ and
$i=85.0^{\circ}$ (Harrop-Allin \etalc\ 1999b, Schwope \etalc\
2001). Table~\ref{tab:freeparams} gives the values of $R_{\mu}$ and the field
orientation parameters for each cycle. We estimate the range on the values of
$\beta$ and $\zeta$ as $\pm10^\circ (1\sigma)$.

The values of the three parameters in Table~\ref{tab:freeparams} are different
from those found by Harrop-Allin~\etalc\ (1999b). We find
$R_{\mu}=0.22\--0.26a$ ($1.3\times 10^{10}$~cm to $1.6\times 10^{10}$~cm),
whereas Harrop-Allin~\etalc\ (1999b) found $R_{\mu}=0.17\--0.20a$. We can
attribute this difference to an improved stream trajectory compared to that of
Harrop-Allin~\etalc\ (1999a). This arises from an incomplete treatment of the
appropriate forces in the frame of reference chosen by Harrop-Allin~\etalc\
(1999a). The difference is small, but for the angles used here the resulting
$R_{\mu}$ is significantly different, with the Harrop-Allin~\etalc\ trajectory
underestimating the values for $R_{\mu}$.

The decrease in the value of $R_{\mu}$ from cycle 29993 to 29995 implies that
the stream penetrates further into the magnetosphere before threading onto the
magnetic field lines (Figure~\ref{fig:streambright}). The effect of the change
in the threading radius can be seen in Figure~\ref{fig:inoutsmooth}. The values
of $\beta\sim 60^{\circ}$, $\zeta\sim 40^{\circ}$ found here are also larger
than those used by Harrop-Allin~\etalc\ (1999b), but consistent within the
uncertainties. Any change in these parameters would imply that the location of
the accretion spot has moved further from the line of centres and the altered
geometry of the magnetic field carries the accretion stream further round the
white dwarf from the line of centres (see Section 5). They are in agreement
with those found by Schwope~\etalc\ (2001) (their figure~6).

\begin{table}
\caption{The values of the coupling radius ($R_{\mu}$) and location of the
accretion spot ($\beta$ and $\zeta$, magnetic longitude and colatitude
respectively) for cycles 29993 and 29995. Values of $\beta$ and $\zeta$ have an
estimated range of $\pm10^\circ (1\sigma)$.}
\begin{center}
\begin{tabular}{lccc}
\hline
Cycle number & $R_{\mu}$ & $\beta$      & $\zeta$\\
\hline
29993        & 0.26$a$   & 55$^{\circ}$ & 30$^{\circ}$ \\ 
29995	     & 0.22$a$   & 60$^{\circ}$ & 40$^{\circ}$ \\
\hline
\end{tabular}
\end{center}
\label{tab:freeparams}
\end{table}

\subsection{Model results}

With the values of $R_{\mu}$ determined above, we can use the model to find the
best fit brightness map, from which we can then determine the brightness per
unit stream length along the stream trajectory. This will show the brightness independent of the actual length of the stream eclipsed in any
particular phase interval. This is important as there are different lengths of
ballistic and magnetically confined stream eclipsed in a given phase interval. 

Figure~\ref{fig:partlc} presents the results of the modelling in two columns
representing the two cycles presented here and the three rows the three energy
bands: red, yellow and blue. We have divided the stream into three sections, as
illustrated in Figure~\ref{fig:streambright}, which enclose the section nearest
the white dwarf, the threading region and the stream in between the two. The
ordinate then represents the total brightness of points eclipsed in these
intervals per unit stream length, that is points in the ballistic and
magnetically confined sections of the stream eclipsed in that phase range. The
results have been calibrated into energy units using the standard star
observation in order to facilitate comparison between the different
bands. However the calibration is only approximate due to a lack of a standard
star on the second night of observations.

\begin{figure}
\centerline{\epsfig{file=partlc.ps,width=6.0cm,angle=-90.0}}
\caption{The above plot shows for cycles 29993 and 29995, in each band, the
total brightness per unit stream length in three sections of the stream. These
intervals include a section of the ballistic trajectory plus those parts of the
stream nearest the white dwarf (section 1), the threading region (section 3)
and the magnetically confined stream between the two (section 2),
c.f. Figure~\ref{fig:streambright}. A brightness enhancement at the threading
region is evident in both cycles, but an enhancement towards the white dwarf is
not seen in the red and yellow bands of cycle 29995.}
\label{fig:partlc}
\end{figure}

\begin{figure}
\centerline{\epsfig{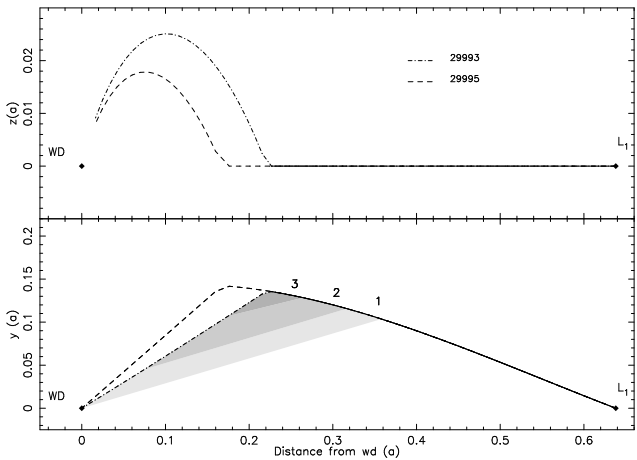}}
\caption{The upper plot shows the model stream points for cycles 29993 and
29995 looking parallel to the orbital plane, while the lower plot shows them
looking down onto the orbital plane. The difference in the stream trajectories
is evident. The shaded regions are three equal phase intervals encompassed by
the ingress of the white dwarf and the final ingress of the threading region
for cycle 29993. These correspond to those parts of the stream used to form
sections 1, 2 and 3 in Figure~\ref{fig:partlc}. (For cycle 29995 the
definitions are equivalent.)}
\label{fig:streambright}
\end{figure}

We can see from Figure~\ref{fig:partlc} that there is enhanced brightness in
the section containing the threading region in all bands in cycle 29993, but
only in the yellow and blue in cycle 29995. The section containing those parts
of the stream nearest the white dwarf are also bright in cycle 29993. However,
in cycle 29995 this is only seen in the blue. The brightness of the different
sections is consistent with the colour ratios from
Section~\ref{sec:colourvariations}. In Figure~\ref{fig:partlc} the threading
region in cycle 29995 is hotter than that in cycle 29993 as the ratio of the
yellow to red is greater for cycle 29995. This is again consistent with the
colour ratios (Section~\ref{sec:colourvariations}). However, the blue to yellow
ratio is similar in the two eclipse ingresses suggesting that the increase has
occurred in both blue and yellow bands.

\section{Discussion}

\subsection{Comparison with earlier results}

The model fits of Harrop-Allin~\etalc\ (1999b) and Harrop-Allin (1999) for one
pole accretion, show a general enhancement in the threading region and towards
the white dwarf. Some eclipses (for example cycle 3723 and 3724 in Harrop-Allin
1999) show significantly greater brightness in the threading region of the
white dwarf than others (such as the immediately preceding cycle 3722).

Enhanced brightness regions are also found by Vrielmann \& Schwope (2001) and
Kube \etalc\ (2000) who apply similar modelling techniques. Vrielmann \&
Schwope (2001) apply their Accretion Stream Mapping technique to H$\beta$,
H$\gamma$ and He\ {\sc ii}\ $\lambda$\ 4686\AA\ observations of HU Aqr. They
find a brightness enhancement at the threading region, but not towards the
white dwarf, similar to cycle 29995, so implying that this may be absent in
their emission lines. Kube \etalc\ (2000) used the C{\sc iv}\ $\lambda$\
1550\AA\ line emission from UZ For, with a 3-dimensional stream model. They
found three regions of enhanced brightness: one on the ballistic accretion
flow, and two on the magnetically confined section. They suggest that the
enhancements on the magnetically confined section are caused by irradiation of
denser sections of the stream with a large area near the accretion region, and
a smaller region near to the threading region. This may be the case in cycle
29993, where we find a bright stream near the white dwarf, and again near the
threading region. Although Kube~\etalc\ find no enhancement actually at the
threading region, this may be a result of an increase in the density of the
stream as it approaches this point, resulting in an increase in the continuum
optical depth, and hence a decrease of the C{\sc iv}\ $\lambda$\ 1550\ \AA\
equivalent width, and so is not necessarily indicative of a faint stream at
this region.

\subsection{Heating in the magnetically confined stream}

Heating of the magnetic section of the stream has been modelled by Ferrario \&
Wehrse (1999) for a stream which is assumed to thread onto the field lines at a
coupling radius $r_{c}$ (our $R_{\mu}$) from the white dwarf, over a radial
distance $\Delta r_{c}$ in the orbital plane. This provides an opportunity for
a comparison between our observationally derived results, and the main points
of their theoretical model results.

Ferrario \& Wehrse (1999) consider two heating mechanisms: irradiation by the
X-ray component from the accretion shock (the soft X-rays being up to $\approx
4000$ times more efficient at heating the stream than the hard X-rays) and the
effects of magnetic reconnection in the stream-field interaction over the
threading region $\Delta r_{c}$. The comparison we make with these theoretical
models is qualitative as the geometrical structure of the magnetically confined
accretion flow in Ferrario \& Wehrse is funnel-shaped, compared to our linear
trajectory. However, our results showing brightness enhancements towards the
threading region are consistent with their models which incorporated magnetic
heating in the threading region. This implies that some magnetic heating
mechanism is needed.

\subsection{Temporal variations in the stream profile}

Based on the colour ratios we deduced that the threading region was brighter
and hotter in cycle 29995 than in cycle 29994
(Section~\ref{sec:colourvariations}). There are therefore significant changes
in the threading region on the timescale of the orbital period (125 mins).

Although we have only modelled two cycles here, we can see that there is a
difference in the brightness confirming the variability seen in the colour
ratios and in the stream eclipse profiles from the light curves (see
Figure~\ref{fig:inoutsmooth}). The large brightness enhancement near the white
dwarf in cycle 29993 could be irradiated material which cools over the
timescale of the next orbital period, leaving a fainter region as seen in cycle
29995 where the observed enhancement is less pronounced. However, without the
model of the intermediate cycle, which we have excluded for reasons discussed
previously, we cannot be certain and it will require more high signal-to-noise
ratio observations of consecutive eclipses to investigate whether this in fact
occurs. What is clear from the model results, and indeed directly from the raw
light curves and colours, is that the emission from the whole stream is highly
dynamic and unstable.

Dramatic changes in the stream eclipse profiles were also observed in the low
state (Harrop-Allin \etalc\ 2001, Glenn \etalc\ 1994). These cycle-to-cycle
changes in stream brightness and trajectory require a revision in our view of
the stream, and in the manner in which the magnetic heating takes place in the
threading region. Future treatments may need to consider large scale magnetic
instabilities, and quasi-cyclic behavior.

\section{Summary and conclusions}

We have carried out high signal-to-noise ratio observations of HU Aqr using
S-Cam2 on the WHT on two nights. The system was in a high accretion state, and
from the single sharp change in the eclipse profile and archive soft X-ray
light curves in this state, we infer that matter was accreting at only one pole
on the white dwarf. At the onset of the eclipse, the accretion stream is the
source of more than half of the optical emission from the system. The system
brightness was similar from orbit to orbit, and the eclipse duration remained
constant, but the shape of the accretion stream eclipse changed significantly,
ending at $\phi=0.979$, $0.977$ and $0.982$.

We find eclipse durations which are unchanged from past optical studies
(Harrop-Allin \etalc\, 1999b, Schwope \etalc\ 1997), but shorter than those
deduced in the soft X-rays by Schwope \etalc\ (2001). However, the location of
the accretion region is similar to that found by Schwope \etalc\ (2001), so
that there is no evidence that the optical emission is from higher latitudes on
the white dwarf than the soft X-rays. This indicates that the duration of the
soft X-ray eclipses may be affected by absorption, as they suggested.

The duration of the egress of the accretion region in the optical is $8$~s,
compared to $1.3$~s in soft X-rays (Schwope \etalc\ 2001). This indicates
clearly that the region emitting cyclotron radiation is extended by a factor of
$\sim 5$ by comparison with the soft X-ray emitting region, which Schwope
\etalc\ (2001) calculated as subtending an angle of $3^{\circ}$.

We have found significant changes in the colour of the accretion stream from
one eclipse to the next. This indicates that the threading region is hottest in
the last of the eclipses by comparison with the previous two (see
Section~\ref{sec:colourvariations}). We have modelled the stream using the
technique of Hakala (1995) and Harrop-Allin \etalc\ (1999a). This finds that
most of the emission originates from two places, the region close to the white
dwarf and in the threading region. By comparison with the models of Ferrario \&
Wehrse (1999) this indicates that magnetic heating is required in the threading
region. The modelling clearly identifies an increase in brightness in the
threading region for both cycles and an enhancement towards the white dwarf for
cycle 29993. From this change in brightness of the heated regions in the model
streams, and the varying stream eclipse profiles, we suggest that the magnetic
heating in the threading region may be unstable. The implications of the highly
variable stream trajectory and brightness profile should be recognised in
future investigations of the stream properties. A systematic study of series of
consecutive eclipses, with the highest possible signal-to-noise ratio, is
required to investigate the characteristics of the magnetic heating and stream
instabilities.

\section*{ACKNOWLEDGMENTS}

We acknowledge the contributions of other members of the Astrophysics Division
of the European Space Agency at ESTEC involved in the optical STJ development
effort, in particular S.~Andersson, D.~Martin, J.~Page, P.~Verhoeve, and
J.~Verveer. We acknowledge the excellent support given to the instrument's
operation at the WHT by the ING staff, in particular P.~Moore and C.R.~Benn.

\end{document}